\documentclass[pre,floatfix,longbibliography,superscriptaddress,twocolumn,final]{revtex4-1}
\usepackage{graphicx}
\usepackage{amsfonts,amsmath}
\usepackage[caption=false]{subfig}
\usepackage[usenames,dvipsnames,svgnames,table]{xcolor}
\usepackage{hyperref}
\hypersetup{
 bookmarks=true, 
 unicode=false, 
 pdftoolbar=true, 
 pdfmenubar=true, 
 pdffitwindow=false, 
 pdfstartview={FitH}, 
 pdftitle={}, 
 pdfauthor={}, 
 pdfcreator={}, 
 pdfproducer={}, 
 pdfkeywords={} {} {}, 
 pdfnewwindow=true, 
 colorlinks=true, 
 linkcolor=red, 
 citecolor=Green, 
 filecolor=magenta, 
 urlcolor=cyan 
}
\usepackage{booktabs,tabularx}
\usepackage{multi row}
\graphicspath{{./Figures/}} 

\AtBeginDocument{
\heavyrulewidth=1.5pt
\lightrulewidth=.05em
\cmidrulewidth=.03em
\belowrulesep=.65ex
\belowbottomsep=0pt
\aboverulesep=.4ex
\abovetopsep=4pt
\cmidrulesep=\doublerulesep
\cmidrulekern=.5em
\defaultaddspace=.5em
}

\newcommand{\be}{\begin{equation}} 
\newcommand{\ee}{\end{equation}} 
\newcommand{\bea}{\begin{eqnarray}}
\newcommand{\eea}{\end{eqnarray}}

\newcommand{\e}{\mathrm{e}}
\renewcommand{\L}{\mathcal{L}}
\renewcommand{\vec}[1]{#1}

\renewcommand{\bf}[1]{\textbf{#1}} 

\newcommand{\f}[2]{\frac{#1}{#2}}

\newcommand{\s}{\sigma}

\newcommand{\al}{\alpha}
\renewcommand{\b}[1]{\bar{#1}}

\begin{document}
\title{Community detection, link prediction, and layer interdependence in multilayer networks}
	\author{Caterina De Bacco} 
	\affiliation{Santa Fe Institute, 1399 Hyde Park Road, Santa Fe, NM 87501, USA}
	\email{cdebacco@santafe.edu, moore@santafe.edu}
\author{Eleanor A. Power} 
	\affiliation{Santa Fe Institute, 1399 Hyde Park Road, Santa Fe, NM 87501, USA}
\author{Daniel B. Larremore} 
	\affiliation{Santa Fe Institute, 1399 Hyde Park Road, Santa Fe, NM 87501, USA}
\author{Cristopher Moore} 
	\affiliation{Santa Fe Institute, 1399 Hyde Park Road, Santa Fe, NM 87501, USA}

\begin{abstract}
Complex systems are often characterized by distinct types of interactions between the same entities.
These can be described as a multilayer network where each layer represents one type of interaction. These layers may be interdependent in complicated ways, revealing different kinds of structure in the network. 
In this work we present a generative model, and an efficient expectation-maximization algorithm, which allows us to perform inference tasks such as community detection and link prediction in this setting. Our model assumes overlapping communities that are common between the layers, while allowing these communities to affect each layer in a different way, including arbitrary mixtures of assortative, disassortative, 
or directed structure. It also gives us a mathematically principled way to define the interdependence between layers, by measuring how much information about one layer helps us predict links in another layer. In particular, this allows us to bundle layers together to compress redundant information, and identify small groups of layers which suffice to predict the remaining layers accurately. We illustrate these findings by analyzing synthetic data and two real multilayer networks, one representing social support relationships among villagers in South India and the other representing shared genetic substrings material between genes of the malaria parasite.
\end{abstract}

\maketitle


\section{Introduction}	

Networks that describe real-world relationships often have different types of links that connect their nodes. For example, links in a social network may represent friendships, marriages, or collaborations, while links in a transportation network may represent planes, trains, or automobiles. Such networks are called multitype or multilayer networks since each type of link can be separated into its own layer, thereby connecting the same set of nodes in multiple ways. However, understanding the large-scale structure of multilayer networks is made difficult by the fact that the patterns of one type of link may be similar to, uncorrelated with, or different from the patterns of another type of link. These differences from layer to layer may exist at the level of individual links, connectivity patterns among groups of nodes, or even the hidden groups themselves to which each node belongs. Therefore, finding community structure in multilayer networks requires simultaneously considering three related problems: (i) the \emph{multilayer community detection problem}, in which we seek a description of the network that divides the nodes according to groups hidden in the link patterns of multiple layers; (ii) the \emph{layer interdependence problem}, in which we seek a description of the relationships between the layers containing different types of links; and (iii) the \emph{link prediction problem}, in which we seek to accurately predict missing link data by making use of all relevant layers of the network.

These three problems are fundamentally intertwined. Multilayer community detection requires knowing which layers have related structure and which layers are unrelated, since redundant information across layers may provide stronger evidence for clear communities than each layer would on its own. However, measuring layer interdependence requires a working definition of interdependence and a method to measure it. The performance of a model or algorithm on the link prediction task provides exactly such a measure: specifically, whether it is possible to integrate information across layers, using links of one type to help predict those of another type. Thus, in settings where data are well-represented as a multilayer network, as is relevant in inferring genetic and protein-protein interactions in cells~\cite{costanzo2010}, characterizing interdependent infrastructures~\cite{rinaldi2001} and understanding the impact of different social ties in human relationships~\cite{verbrugge1979,wasserman1994}, we desire a model capable of recognizing interdependencies between layers, integrating and merging information between them, and using this information both to classify nodes and to predict links.

In this paper we propose an approach that performs all three of these tasks. We define a generative model for multilayer community detection that is applicable to both directed and undirected networks, as well as networks with integer-weighted links. We provide a highly-scalable expectation-maximization algorithm that fits this model to data, taking as its input a multilayer network and yielding a ``mixed-membership'' partition: nodes are divided into communities or groups, but each node may belong to some extent to multiple groups. While this partition is shared by all the layers, the model allows for different connectivity patterns in each layer, including arbitrary mixtures of assortative, disassortative, and directed structure. 

In addition to classifying nodes, our model also makes link prediction possible: given an incomplete dataset where not all links are known, it assigns probabilities to each pair of nodes that they have an unobserved link of each type. Finally, by sequentially fitting the model to single layers and multiple layers, we show how to determine whether the information provided by additional layers improves link prediction performance, thereby quantifying layer interdependence. We show how to use this method to identify which layers of the network are redundant, and which provide independent information. For instance, we can identify small sets of layers which together capture most of the information about the network. This may be useful in contexts where gathering layers requires an investment of limited resources in the laboratory or the field---for instance, if a social scientist can ask their subjects a limited number of questions, identifying a limited number of types of social relationships. 

Community detection is a fundamental element of network science, yet most community detection algorithms have been developed for single-layer networks. We can use these methods to analyze multilayer networks, either by aggregating all layers to create a single-layer network~\cite{taylor2016,taylor2016enhanced} or by analyzing each layer independently. Multilayer-specific methods that maximize community quality functions such as modularity~\cite{mucha2010} provide non-overlapping partitions, but inherit the issues of their single-layer counterparts, namely a dependence on proper choice of a null model~\cite{bazzi2016,paul2016}. Moreover, these methods do not typically provide a framework to perform link prediction, since without additional assumptions they do not assign probabilities to the presence or absence of a link.

More recent methods for multilayer community detection are based on fitting various generative models via Bayesian inference or maximum likelihood estimation~\cite{paul2015,peixoto2015,schein2014,schein2015,schein2016bayesian,valles2016multilayer}. Our algorithm falls into this category, but differs from most of these models by not assuming \textit{a priori} any specific network structure. Many network models assume \emph{assortative} or \emph{homophilic} community structure, meaning that nodes are more likely to be connected to others in the same community. 
This assumption is often incorrect in food webs, technological networks, and social networks where links consist of nominations, for instance, of a trustworthy or powerful person that they might ask for advice, help or work. Our model avoids this assumption: while it looks for an (overlapping) community structure that is consistent with every layer, it deals happily with networks where some layers are assortative, others are disassortative, yet others have core-periphery or directed structure, and so on. In the case of directed links, it also recognizes that nodes might play different roles, and thus effectively belong to different groups, when forming incoming or outgoing links. 

By measuring the extent to which one layer helps us predict links in another layer, our method also gives a mathematically principled way to measure the relationships between the layers of a multilayer network, including identifying layers which are redundant with others or highly independent from them. This problem arises even in naive algorithms that aggregate layers into a single-layer network. For instance, the multilayer versions of eigenvector centrality or modularity~\cite{mucha2010,sola2013} use weighted averages over layers, requiring them to infer or choose a weight for each layer. A number of recent works have taken a compression approach, aggregating layers with similar structure~\cite{de2015,stanley2015}; in particular, Ref.~\cite{stanley2015} uses a generative model that jointly assigns community memberships to nodes and groups of layers, which they call \emph{strata}. This is similar in spirit to our approach, but our model handles overlapping communities, directed networks, and weighted networks in a unified way. 

In Section~\ref{sec:model} we describe our model and in Section~\ref{sec:em} give an efficient algorithm that fits its parameters to network data. In Section \ref{sec:synthetic}, we provide performance results on synthetic benchmarks and compare with other algorithms, and in Section \ref{sec:linkpr} we discuss how to perform link prediction and measure layer interdependencies. We then apply these concepts to two real world networks, drawn from anthropology and biology, and discuss the results in Section \ref{sec:results} before concluding.

\section{The multilayer mixed-membership stochastic block model}
\label{sec:model}

In this section we describe our model and fix our notation. The network consists of $N$ nodes and has $L$ layers. Each layer has an adjacency matrix $A^{(\alpha)}$, where $A_{ij}^{(\alpha)}$ is the number of edges from $i$ to $j$ of type $\alpha$; alternatively, we can think of $A$ as an $N \times N \times L$ tensor. Our model generates these networks probabilistically, assuming an underlying structure consisting of $K$ overlapping groups.

Each node belongs to each group to an extent described by a $K$-dimensional vector. Since we are interested in directed networks, we give each node $i$ two membership vectors, $\vec{u}_i$ and $\vec{v}_i$, which determine how $i$ forms outgoing and incoming links respectively. (When modeling undirected networks, we set $\vec{u}=\vec{v}$.) Each layer $\alpha$ has a $K \times K$ affinity matrix $w^{(\alpha)}$ describing the density of edges between each pair of groups. The expected number of edges in layer $\alpha$ from $i$ to $j$ is then given by a bilinear form, 
\be
	M_{ij}^{(\al)} = \sum_{k,\ell=1}^K u_{ik} \,v_{j\ell} \,w_{k\ell}^{(\al)} \, .
\ee
Finally, for each $i, j$ and $\alpha$, we choose $A_{ij}^{(\alpha)}$ independently from the Poisson distribution with mean $M_{ij}^{(\alpha)}$. 

Note that while we assume that the membership vectors have nonnegative entries, we do not normalize them. This allows us to account easily for heterogeneous degree distributions, since multiplying $\vec{u}_i$ or $\vec{v}_i$ by a constant increases the expected out- or in-degree without changing the distribution of neighbors to which a given one of $i$'s edges connects.

Note also that while this model supposes that nodes have the same group membership in all layers, it allows the structure of each layer to vary arbitrarily with respect to these groups. For instance, some layers could be assortative and others disassortative, with affinity matrices $w^{(\alpha)}$ which are large on or off the diagonal; other layers could have strongly directed structure, with asymmetric $w^{(\alpha)}$, or core-periphery structure, where $w^{(\alpha)}$ has one large entry on the diagonal. 

For a single layer, our model is similar to existing mixed-membership block models~\cite{newman2007,airoldi2008,ramasco2008,ball2011,gopalan2013,yang2014,zhou2015}. Some of these use mixtures of Bernoulli random variables; we follow~\cite{ball2011} in using the Poisson distribution since it leads to a tractable and efficient expectation-maximization algorithm. The Poisson distribution also allows us to model multigraphs or integer-weighted networks. However, in our applications here we will focus on the sparse case where $M_{ij}^{(\alpha)}$ is small, and assume for simplicity that $A_{ij}^{(\alpha)}$ is $0$ or $1$. 

Our model also bears a close mathematical relationship to topic models~\cite{PLSA,LDA}, which generate bipartite weighted graphs of documents and words based on their relevance to mixtures of topics. More generally, it can be viewed as a variant of non-negative tensor factorization (see e.g.~\cite{Kolda2009} for a review) and in particular of Poisson tensor factorization~\cite{lee1999,canny2004,dunson2005,cemgil2009,gopalan-rec,schein2014,schein2015,chi2012tensors}. However, the affinity matrices $w^{(\alpha)}$, which allow different layers to be assortative or disassortative, correspond to a kind of Tucker decomposition~\cite{TUCKER2} of tensor rank $K^2$. This is more general than the PARAFAC/CANDECOMP decomposition~\cite{PARAFAC,CANDECOMP}, which corresponds to the special case of our model where $w^{(\alpha)}$ is diagonal for each $\alpha$. In the undirected case where $\vec{u}=\vec{v}$, PARAFAC thus assumes a purely assortative structure, where a link between two nodes can only exist if their membership vectors overlap.

Various kinds of Poisson Tucker decomposition for dynamic and multilayer networks have also been proposed very recently in the machine learning community, particularly in~\cite{schein2014,schein2015,schein2016bayesian}. Indeed, our model is very nearly a special case of that of~\cite{schein2016bayesian}, which has additional parameters intended to model datasets with both multiple types of links and multiple time steps. The main difference between these works and our approach is that they impose priors on the ``core tensor'' parameters [analogous to $w^{(\alpha)}_{k\ell}$] and they use Monte Carlo sampling for Bayesian inference. In contrast, we find point estimates of these parameters using an expectation-maximization algorithm, detailed in the next section. Below we compare our results to the algorithm of~\cite{schein2015}, which is designed for the same kinds of datasets as ours.

\section{The Expectation-Maximization algorithm}
\label{sec:em} 

Given an observed multilayer network with adjacency tensor $A$, our goal is to simultaneously infer the nodes' membership vectors and the affinity matrices for each layer. In this section, we describe an efficient algorithm which does this by the method of maximum likelihood. 

Let $\Theta$ be shorthand for all $2NK+K^2 L$ model parameters, i.e., the $u_{ik}$, $v_{i\ell}$, and $w_{k\ell}^{(\alpha)}$. Assuming that all $\Theta$ are equally likely \textit{a priori}, the probability of $\Theta$ given $A$ is proportional to the probability of $A$ given $\Theta$. Using the Poisson distribution function gives 
\be
	P(\Theta \mid A) \propto P(A \mid \Theta) 
	= \prod_{i,j=1}^{N} \prod_{\al=1}^L \frac{\e^{-M_{ij}^{(\al)}} (M_{ij}^{(\al)})^{A_{ij}^{(\al)}}}{A_{ij}^{\al}!}
	\label{eq:likelihood}
\ee
(One could also impose a prior $P(\Theta)$ and perform maximum \textit{a posteriori} inference~\cite{dunson2005,cemgil2009,schein2015}, but we have not done this here.) The log-likelihood is then 
\be
	\L(\Theta) 
	 = \sum_{i,j,\alpha}
	\left[ A_{ij}^{(\al)} \log \sum_{k\ell} u_{ik} v_{j\ell} w_{k\ell}^{(\al)}
	 - \sum_{k\ell} u_{ik} v_{j\ell} w_{k\ell}^{(\al)} \right]
	 \label{logL}
\ee
where we omit the terms $\log A_{ij}^{(\al)}!$ since they depend only on the data.

We wish to find the $\Theta$ that maximizes Eq.~\eqref{logL}. This is computationally difficult, but we can make it more tractable with a classic variational approach. For each $i,j,\alpha$ with $A_{ij}^{(\alpha)} = 1$, consider a probability distribution $\rho_{ijk\ell}^{(\alpha)}$ over pairs of groups $k,\ell$: this is our estimate of the probability that that edge exists due to $i$ and $j$ belonging to groups $k$ and $\ell$ respectively. (If the network is a multigraph and $i,j$ have multiple links of the same type, we give each one its own distribution $\rho$; below we assume for simplicity that this does not occur.) Jensen's inequality $\log \overline{x} \ge \overline{\log x}$ then gives
\begin{align}
\log \sum_{k\ell} &u_{ik} v_{j\ell} w_{k\ell}^{(\al)} 
= \log \sum_{k\ell} \rho_{ijk\ell}^{(\alpha)} \frac{u_{ik} v_{j\ell} w_{k\ell}^{(\al)}}{\rho_{ijk\ell}^{(\alpha)}} \nonumber \\
&\ge \sum_{k\ell} \rho_{ijk\ell}^{(\alpha)} \log \frac{u_{ik} v_{j\ell} w_{k\ell}^{(\al)}}{\rho_{ijk\ell}^{(\alpha)}} \nonumber \\
&= \sum_{k\ell} \rho_{ijk\ell}^{(\alpha)} \log u_{ik} v_{j\ell} w_{k\ell}^{(\al)} 
- \sum_{k\ell} \rho_{ijk\ell}^{(\alpha)} \log \rho_{ijk\ell}^{(\alpha)} \, . 
\label{eq:variational}
\end{align}

Moreover, this holds with equality when
\begin{equation}
\label{eq:boltzmann}
	\rho_{ijk\ell}^{(\al)} = \frac{u_{ik} v_{j\ell} w_{k\ell}^{(\al)} }{\sum_{k'\ell'} u_{ik'} v_{j\ell'} w_{k'\ell'}^{(\al)} } \, . 
\end{equation}
Thus maximizing $\L(\Theta)$ is equivalent to maximizing 
\begin{align}
\L(\Theta,\rho)
&= \sum_{i,j,\alpha,k,\ell}
	\Big[ A_{ij}^{(\al)} 
	\left( \rho_{ijk\ell}^{(\alpha)} \log u_{ik} v_{j\ell} w_{k\ell}^{(\al)} 
- \rho_{ijk\ell}^{(\alpha)} \log \rho_{ijk\ell}^{(\alpha)} \right) \nonumber \\
	& \quad - u_{ik} v_{j\ell} w_{k\ell}^{(\al)} \Big] 
\end{align}
with respect to both $\Theta$ and $\rho$. 

The expert reader will recognize that this variational argument is simply classical thermodynamics in disguise. Fix the parameters $\Theta$ and consider a spin system where each edge, i.e., each triple $(i,j,\alpha)$ with $A_{ij}^{(\alpha)} = 1$, has a state consisting of a pair of groups $(k,\ell)$. Define the Hamiltonian as 
\begin{align*}
H &= - \sum_{(i,j,\alpha): A_{ij}^{(\alpha)}=1} \log u_{ik(i,j,\alpha)} v_{j\ell(i,j,\alpha)} w_{k(i,j,\alpha),\ell(i,j,\alpha)}^{(\alpha)} \\
& \quad - \sum_{i,j,\alpha,k',\ell'} u_{ik'} v_{j\ell'} w_{k'\ell'}^{(\al)} 
\end{align*}
(note that the second term is constant). Then the Boltzmann distribution is a product distribution of the distributions $\rho_{ij}^{(\alpha)}$ given by Eq.~\eqref{eq:boltzmann} on each edge. Moreover, $-\L(\Theta,\rho)$ is the free energy $E-TS$ where $T=1$, and we recover the familiar fact that this is minimized by the Boltzmann distribution. In this context, finding the maximum-likelihood estimate of the parameters $\Theta$ corresponds to minimizing the free energy of this spin system.

We can maximize $\L(\Theta,\rho)$ by alternatively updating $\rho$ and $\Theta$. This general approach is called an expectation-maximization (EM) algorithm: the expectation step computes the marginals of the Boltzmann distribution for the current estimate of the parameters, and the maximization step finds the most-likely value of the parameters given those marginals. The fact that the Boltzmann distribution takes a simple product form makes the expectation step especially simple, making the algorithm highly efficient.

The update equations for $\Theta$ in the maximization step can be can be derived by computing the partial derivative of $\L(\Theta,\rho)$ with respect to the various parameters. For instance,
\be
\frac{\partial \L(A,\rho)}{\partial u_{ik}} 
= \sum_{j,\ell,\al} \left[ \frac{A_{ij}^{(\al)}\, \rho_{ijk\ell}^{(\al)}}{u_{ik}} - v_{j\ell} w_{k\ell}^{(\al)} \right] \, . 
\ee 
Setting this to zero, and doing the same for the partial derivatives with respect to $v_{j\ell}$ and $w_{k\ell}^{(\al)}$, gives
\begin{gather}
{u}_{ik}
= \frac{\sum_{j,\al} A_{ij}^{(\al)} \sum_\ell \rho_{ijk\ell}^{(\al)}}
{\sum_\ell \left( \sum_{j} v_{j\ell} \right) \left( \sum_{\al} w_{k\ell}^{(\al)} \right)}
\label{uik_dis} \\
{v}_{j\ell}
= \frac{\sum_{i,\al} A_{ij}^{(\al)} \sum_k \rho_{ijk\ell}^{(\al)}}
{\sum_k \left( \sum_i u_{ik} \right) \left( \sum_{\al} w_{k\ell}^{(\al)} \right)}
\label{vjk_dis} \\
{w}_{k\ell}^{(\al)}= \f{\sum_{ij}A_{ij}^{(\al)} \rho_{ijk\ell}^{(\al)}}{\left(\sum_i u_{ik} \right) \left( \sum_j v_{j\ell} \right)}
\label{omegaka_dis} \, .
\end{gather}

The EM algorithm thus consists of randomly initializing the parameters $\Theta$, and then repeatedly alternating between updating $\rho$ using Eq.~\eqref{eq:boltzmann} and updating $\Theta$ using Eqs.~\eqref{uik_dis}-\eqref{omegaka_dis} until it reaches a fixed point. This fixed point is a local maximum of $\L(\Theta,\rho)$, but it is not guaranteed to be the global maximum. Therefore, we perform multiple runs of the algorithm with different random initializations for $\Theta$, taking the fixed point with the largest value of $\L(\Theta,\rho)$. The computational complexity per iteration scales as $O(MK^2)$ where $M$ is the total number of edges summed over all layers and $K$ is the number of groups. In practice we find that our algorithm converges within a fairly small number of iterations. Thus it is highly scalable, with a total running time roughly linear in the size of the dataset.

Once we converge to a fixed point, we can assign nodes to communities by normalizing the membership vectors to $\b{u}_i = u_i / \sum_k u_{ik}$, so that for each $i$ we have $\sum_k \b{u}_{ik} = 1$. This approach was used in~\cite{ball2011} as an method for classifying nodes in overlapping communities; however, since we allow $\vec{u}_i$ and $\vec{v}_i$ to be distinct, the ``outgoing'' and ``incoming'' assignments of a node might differ. These are \emph{soft} assignments, meaning that nodes can belong to more than one community. If one wishes to obtain a \emph{hard} assignment, one can assign each node to the single community corresponding to the maximum entry of $\vec{u}$ or $\vec{v}$, but the overlapping character of the community structure is then lost. 

We call our model and its associated algorithm 
 \mbox{{\small MULTITENSOR}}. A numerical implementation is available for use under an open source license \footnote{Code available at \href{https://github.com/cdebacco/{\small MULTITENSOR}}{github.com/cdebacco/{\small MULTITENSOR}}}.

\section{Results on synthetic networks}
\label{sec:synthetic}

We tested {\small {\small MULTITENSOR}}'s ability to detect community structure synthetic networks using the multilayer benchmark proposed in Ref.~\cite{bazzi2016}. This model is somewhat different from ours: rather than having mixed-membership vectors that are the same in every layer, they use hard partitions where each node belongs to a single group, but they allow these partitions to vary from layer to layer. The partitions in different layers are correlated by a so-called layer interdependence tensor. For simplicity, we use a one-parameter version of this tensor with dependency $p \in [0,1]$: when $p=0$ the partitions between layers are independent, and when $p=1$ they are identical. Once these partitions are chosen, they generate edges in each layer according to a degree-corrected block model~\cite{karrer2011}, with a user-specified degree distribution and affinity matrix.

We used this benchmark to generate synthetic networks with $N=300$ nodes, $L=4$ layers, and $K=5$ communities. For each layer we used a truncated power-law degree distribution with exponent $\gamma =-3$, minimum degree $k_{min}=3$ and maximum degree $k_{max}=30$. We varied the affinity matrix of the block model according to a mixing parameter $\mu$: if $\mu=0$ all edges lie within communities, and if $\mu=1$ edges are assigned regardless of the community structure. 

We define our algorithm's accuracy on these benchmarks in terms of how well the inferred membership vectors match the ground-truth distribution of memberships across layers. That is, we hope that, after normalizing so that $\sum_k \b{u}_{ik}=1$, each $u_{ik}$ is close to the fraction of layers in which node $i$ belongs to group $k$, which we denote $\b{u}^0_{ik}$. We quantify the similarity between these two distributions using two measures. The first is their inner product, often called the \emph{cosine similarity}  (CS), averaged over all the nodes:
\be\label{CS}
	\mathrm{\mathrm{CS}} 
	= \frac{1}{N} \sum_{i=1}^{N} \frac{\b{u}^0_i \cdot \b{u}_i}{| \b{u}^0_i | | \b{u}_i |} \, ,
\ee
where in the denominator $| \b{u} |$ denotes the Euclidean norm. Here $\mathrm{CS}=1$ corresponds to perfect accuracy. The second is the $L_1$ error between the two distributions, also known as their statistical distance or total variation distance, averaged over all the nodes:
\be\label{L1}
	L_1 = \frac{1}{2N} \sum_{i=1}^{N} \left\| \b{u}^0_i - \b{u}_i \right\|_1 
	= \frac{1}{2N} \sum_{i=1}^{N} \sum_{k=1}^{K} \left| \b{u}^0_{ik} - \b{u}_{ik} \right| \, .
\ee
The factor of $1/2$ is used so that this distance ranges from $0$ for identical distributions to $1$ for distributions with disjoint support. In both measures, we give ourselves the freedom to permute the groups, so that the inferred groups of our model correspond to the groups of the benchmark. Thus we maximize the cosine similarity $\mathrm{CS}$, and minimize the $L_1$ error, over all $K!$ permutations of the $K$ groups.

For comparison, we use two other algorithms that infer overlapping multilayer partitions. 
The first is the restricted diagonal version of our model, which only allows diagonal affinity matrices $w^{(\al)}$; as discussed above, this is equivalent to the Poisson version of PARAFAC tensor factorization~\cite{chi2012tensors} and in the undirected case $\vec{u}=\vec{v}$ this corresponds to assume an assortative network structure. The second algorithm is a fully Bayesian Poisson tensor factorization (BPTF)~\cite{schein2015}. The main differences between these two models are the prior information and the optimization approach. The former considers a uniform prior and calculates point estimates of the parameters using an iterative algorithm similar to ours. The BPTF algorithm instead assumes Gamma-distributed parameters and updates the parameters of these distributions instead of the point estimates; in the end it uses the geometric mean of these distributions as its estimate of the parameters.

\begin{table*}[ht]

 \begin{tabular}{lccccccccccccccccccccccc}
	\toprule
\multirow{2}{4em}{$\mu=0.0$ } & \multicolumn{4}{c}{$p=0.5$} && \multicolumn{4}{c}{$p=0.8$} && \multicolumn{4}{c}{$p=0.9$} \\ 
 \cmidrule(l){2-5} \cmidrule(l){7-10} \cmidrule(l){12-15} 
& CS & $\s$ &$L_{1}$ &$\s$ && CS & $\s$ &$L_{1}$ &$\s$ && CS & $\s$ &$L_{1}$ &$\s$ \\ \hline
{\footnotesize MULTITENSOR} & \bf{ 0.66} & 0.06 & 0.58 & 0.07 && \bf{ 0.93} & 0.07 & \bf{ 0.19} & 0.09 && \bf{ 0.99} & 0.01 & \bf{ 0.07} & 0.01 \\ 
Diagonal & 0.65 & 0.05 & 0.60 & 0.06 && 0.84 & 0.06 & 0.31 & 0.08 && 0.97 & 0.04 & 0.10 & 0.05\\ 
BPTF & \bf{ 0.66} & 0.06 & \bf{ 0.57} & 0.06 && 0.89 & 0.07 & 0.24 & 0.08 && 0.96 & 0.03 & 0.10 & 0.04 \\ 
 	\end{tabular}
	
 \vspace{0.01in}
 
 \begin{tabular}{lccccccccccccccccccccccc} 
	\toprule
\multirow{2}{4em}{$\mu=0.1$ } & \multicolumn{4}{c}{$p=0.5$} && \multicolumn{4}{c}{$p=0.8$} && \multicolumn{4}{c}{$p=0.9$} \\ 
 \cmidrule(l){2-5} \cmidrule(l){7-10} \cmidrule(l){12-15} 
& CS & $\s$ &$L_{1}$ &$\s$ && CS & $\s$ &$L_{1}$ &$\s$ && CS & $\s$ &$L_{1}$ &$\s$ \\ \hline
{\footnotesize MULTITENSOR} &\bf{ 0.63} & 0.05 & 0.62 & 0.05 && \bf{ 0.92} & 0.06 & \bf{ 0.22} & 0.08 && \bf{ 0.98} & 0.01 & \bf{ 0.11} & 0.01 \\ 
Diagonal & \bf{ 0.63} & 0.04 & 0.62 & 0.04 && 0.87 & 0.06 & 0.29 & 0.07 && 0.94 & 0.07 & 0.17 & 0.08\\ 
BPTF & 0.62 & 0.06 &\bf{ 0.61} & 0.05 && 0.84 & 0.07 & 0.32 & 0.08 && 0.93 & 0.06 & 0.18 & 0.07 \\ 
 	\end{tabular}
	
 \vspace{0.01in}
 
 \begin{tabular}{lccccccccccccccccccccccc}
	\toprule
\multirow{2}{4em}{$\mu=0.5$ } & \multicolumn{4}{c}{$p=0.5$} && \multicolumn{4}{c}{$p=0.8$} && \multicolumn{4}{c}{$p=0.9$} \\ 
 \cmidrule(l){2-5} \cmidrule(l){7-10} \cmidrule(l){12-15} 
& CS & $\s$ &$L_{1}$ &$\s$ && CS & $\s$ &$L_{1}$ &$\s$ && CS & $\s$ &$L_{1}$ &$\s$ \\ \hline
{\footnotesize MULTITENSOR} &\bf{ 0.55} & 0.03 & 0.70 & 0.02 && \bf{ 0.55} & 0.05 & \bf{ 0.67} & 0.04 && \bf{ 0.59} & 0.07 & 0.59 & 0.07 \\ 
Diagonal & \bf{ 0.55} & 0.03 & 0.71 & 0.02 && 0.53 & 0.03 & 0.69 & 0.03 && 0.58 & 0.07 & 0.63 & 0.06 \\ 
BPTF & 0.52 & 0.03 & \bf{ 0.69 }& 0.03 && 0.49 & 0.05 & 0.68 & 0.04 && \bf{ 0.59} & 0.07 & \bf{ 0.58} & 0.06 \\ \bottomrule
 	\end{tabular}
	\caption{Performance in detecting overlapping partitions on synthetic networks, using our {\small MULTITENSOR} algorithm, the diagonal special case which corresponds to the Poisson version of PARAFAC tensor decomposition~\cite{chi2012tensors}, and Bayesian Poisson Tensor Factorization~\cite{schein2015}. Benchmark networks were generated with the model of~\cite{bazzi2016} with interdependence $p$ and mixing parameter $\mu$: the community structure is stronger when $p$ is large and $\mu$ is small. For each $(\mu,p)$ pair we measure the cosine similarity (CS) and the average $L_1$ error between the planted and inferred structure averaged over $50$ independently generated benchmark networks. For each network, we run each network $50$ times with independently random initializations, and use the parameters given by the highest-likelihood fixed point. Good performance corresponds to high cosine similarly and low $L_1$ error, and the best performance for each pair of parameter values is indicated by boldface. The errors $\s$ are the standard deviation over the 50 benchmarks.}
\label{table:sim} 
\end{table*}

In Table~\ref{table:sim} we report the best results in terms of cosine similarity and $L_1$ error obtained by three algorithms: {\small MULTITENSOR}, its diagonal special case (or Poisson PARAFAC), and BPTF. By varying the layer interdependence $p$ and the mixing parameter $\mu$, we range over cases where the community structure is relatively easy to infer to those where it is much harder. Specifically, inference is easier when $p$ is large, so that the layers are strongly correlated, and $\mu$ is small, so that most links are within communities. For each pair $(\mu,p)$ we generated $50$ independent benchmark networks, and for each network and each algorithm we performed 50 independent runs with independently random initial conditions, taking the fixed point with highest likelihood. We defined convergence numerically by testing whether $\L(\Theta,\rho)$ has not improved by more than $0.1$ for $10$ iterations.

In every case our algorithm achieves the highest cosine similarity, and in the majority of the cases the smallest $L_1$ error, indicated in boldface. All three algorithms perform poorly in the hard regime where one of the two parameters introduces a high level of stochasticity in the network, i.e. when either $\mu=0.5$ or $p=0.5$. In the other cases our algorithm is significantly better according to both measures.

The benchmarks of~\cite{bazzi2016} are assortative in every layer, and the diagonal and BPTF algorithms work fairly well. To illustrate the greater flexibility of our algorithm, we also generated synthetic networks with different kinds of structure in different layers.. Specifically, we generated layers whose structures are 
assortative ($w_{11} = w_{22} > w_{12} = w_{21}$), 
disassortative ($w_{11} = w_{22} < w_{12} = w_{21}$), 
core-periphery ($w_{11} > w_{12} = w_{21} > w_{22}$), 
and directed with a bias from the first group to the second one ($w_{12} > w_{11} = w_{22} > w_{21}$).

We considered three types of networks, all having $K=2$ groups and $N=300$ nodes, but with different numbers and kinds of layers (see Table \ref{table:AucSynthetic}). In each one the groups are of equal size, with un-mixed group memberships (i.e., $u_i = v_i = (0,1)$ or $(1,0)$). The first type of network has $L=2$ layers, one assortative and one disassortative; the second network has $L=4$, with two assortative and two disassortative layers; the last one has $L=4$, with one layer each with assortative, disassortative, core-periphery, and biased directed structure. 

We generated $10$ independent samples of each of these types of network and calculated the CS and the $L_{1}$ norm between the inferred membership and the ground truth using the maximum-likelihood fixed point over $10$ runs of each algorithm with different random initial conditions. As shown in Table \ref{table:synCSL1}, {\small MULTITENSOR} achieves significantly greater performance than the diagonal or BPTF algorithms in all three cases, due to its flexibility in modeling mixtures of these different types of structure. 


\begin{table}[htbp]
\resizebox{1.0\linewidth}{!}{
\begin{tabular}{lccccc} \toprule
 G & K & L & $\langle E \rangle$ & Structure \\ \hline
1 & 2 & 2 & 1980 & \{assort, disassort\} \\
2 & 2 & 4 & 3960,2250 & \{assort, disassort, core-per, dir. disassort\} \\
3 & 2 & 4 & 1980 & \{2 assort, 2 disassort\} \\ \bottomrule
\end{tabular}
}
\caption{ Description of synthetic network structures. $\langle E \rangle$ is the average number of edges per layer, all networks have $N=300$ nodes. For networks $G=1,3$ we used affinity matrices $W^{a}$ and $W^{d}$ for the assortative and disassortative layers respectively, with entries $w_{11}^{a}=w_{22}^{a}=w_{12}^{d}=w_{21}^{d}= 0.04$, $w_{12}^{a}=w_{21}^{a}=w_{11}^{d}=w_{22}^{d}=0.004$ so that $\langle E \rangle=1980$; for $G=2$ the first two layers (1 assortative and 1 disassortative) have affinity matrices $W^{a}$ and $W^{b}$ with entries $w_{11}^{a}=w_{22}^{a}=w_{12}^{b}=w_{21}^{b}= 0.08$, $w_{12}^{a}=w_{21}^{a}=w_{11}^{b}=w_{22}^{b}=0.008$ so that $\langle E \rangle=3960$; the third and fourth (1 core-periphery and 1 directed disassortative) have affinity matrices $W^{c}$ and $W^{d}$ with entries $w_{11}^{c}=w_{12}^{d}=0.08$, $w_{12}^{c}=w_{21}^{c}=w_{11}^{d}=w_{22}^{d}=0.008$ and $w_{22}^{c}=w_{21}^{d}=0.004$ so that $\langle E \rangle=2250$.}
\label{table:AucSynthetic}
\end{table}

\begin{table*}[ht]
 \begin{tabular}{lcccccccccccccccccccccccc}
	\toprule
 \multirow{2}{4em}{ } & \multicolumn{4}{c}{$G=1$} && \multicolumn{4}{c}{$G=2$} && \multicolumn{4}{c}{$G=3$} \\ 
 \cmidrule(l){2-5} \cmidrule(l){7-10} \cmidrule(l){12-15} 
& CS & $\s$ &$L_{1}$ &$\s$ && CS & $\s$ &$L_{1}$ &$\s$ &&CS & $\s$ &$L_{1}$ &$\s$ \\ \hline
{\footnotesize MULTITENSOR} & \textbf{0.984} & 0.008 & \textbf{0.06} & 0.01 && \textbf{0.990} & 0.001 & \textbf{0.058} & 0.003 && \textbf{0.989} & 0.001 & \textbf{0.056} & 0.002 \\
Diagonal & 0.61 & 0.01 & 0.49 & 0.01 && 0.91 & 0.01 & 0.22 & 0.01 && 0.63 & 0.02 & 0.48 & 0.01 \\
BPTF & 0.60 & 0.02 & 0.48 & 0.01&& 0.90 & 0.03 & 0.21 & 0.05 && 0.63 & 0.03 & 0.48 & 0.01 \\ \bottomrule
 	\end{tabular}
	\caption{Cosine similarity and $L_{1}$ norm for mixed structure synthetic networks. CS as in Eq. (\ref{CS}) and $L_{1}$ norm as in Eq. (\ref{L1}) are calculated between membership vectors $\b{u}$, $\b{v}$ of the inferred partitions and the ground truth [$\b{u}_{0}=\b{v}_{0}=(1,0),\,(0,1)$ for nodes in group 1 and 2, respectively]. Results are averages and standard deviations of the results on the two memberships over 10 networks sampled from each network type $G$, as described in Table \ref{table:AucSynthetic}. The best performance is indicated in boldface. } 
\label{table:synCSL1} 
\end{table*}

\section{Learning layer interdependence via link prediction}
\label{sec:linkpr}

Despite the fact that a network may have multiple layers, there is no guarantee that the structure of one layer is related to the structure of any other. In fact, depending on the context, it may even be desirable that two layers are entirely uncorrelated, since then they reveal different kinds of information about the underlying structure. The \emph{layer interdependence problem} consists of identifying which sets of layers are structurally related, and quantifying the strengths of those interrelationships. 

We are not the first to provide a solution to the layer interdependence problem. One intuitive approach is to independently infer the community structure of each layer and then simply compute pairwise correlations between the community partitions of each layer~\cite{larremore2013}. While this approach is straightforward, it is unable to use the information in some layers to assist with inference of structure in other layers, since every layer is treated independently. Another approach has been to cluster the inferred affinity matrices, analogous to our $w^{(\al)}$, during the parameter inference and estimation procedures, thus gathering layers into ``strata''~\cite{stanley2015}. 

However, while we also explore clustering the $w^{(\al)}$ below, neither of these methods captures the general kind of independence we are interested in. For instance, if two layers use the same group labels for the vertices, but one layer is assortative while the other is disassortative, then knowing one of them is very helpful in predicting the other. In this sense they are closely related, even though they are statistically very different, and indeed anticorrelated with each other.

Our proposal to capture this more general kind of interdependence is based on the idea that two layers are interdependent if and only if the structure of one layer provides meaningful knowledge about the structure of the other. Specifically, by performing link prediction in one layer with or without information about another layer, we quantify the extent to which these two layers are related. 

Since our model is generative, it naturally includes a framework to predict links given partially-observed data: simply use the known data to estimate the parameters, and then use these estimated parameters to compute $M_{ij}^{(\al)}$, i.e., the expected number of links of type $\alpha$ between each pair of nodes $i,j$. We then rank each missing entry in the adjacency tensor according to $M_{ij}^{(\al)}$. The method succeeds to the extent that true missing links are given a higher estimate of $M_{ij}^{(\al)}$ than false ones. We follow~\cite{Clauset2008} in defining the accuracy as the area under the receiver-operator curve or AUC~\cite{hanley1982}. This is the probability that a random true positive is ranked above a random true negative; thus the AUC is $1$ for perfect prediction, and $1/2$ for chance.

To test our ability to predict layer $\alpha$, we perform experiments with $5$-fold cross validation. That is, we hold out $20\%$ of its adjacency matrix $A^{(\al)}$, hiding those entries from the algorithm. We infer the model parameters using the remaining $80\%$ as a training dataset, with or without knowledge of other layers of the network, and measure the inferred model's accuracy on the held-out entries in $A^{(\alpha)}$. 
The independence of the training and test datasets makes cross-validation a robust method against overfitting.
Note that holding out $20\%$ of a layer does not mean removing 20\% of the nodes or 20\% of the links, but rather hiding 20\% of the entries of its adjacency matrix, including both zeros and ones. This means that we just the accuracy of our link prediction algorithm on both links and non-links. 

The final AUC is the average obtained over the $5$ folds, each of which holds out a different subset of $20\%$. Clearly this AUC depends both on the layer $\alpha$ we are trying to predict, and on what set of other layers we give the algorithm access to.

Given this framework, we can define the pairwise interdependence between two layers $\alpha, \beta$ as follows. We perform link prediction on layer $\alpha$, with $20\%$ of the entries of $A^{(\alpha)}$ hidden; but we do this first by giving the algorithm access only to the rest of $A^{(\alpha)}$, and then by giving it access to all of $A^{(\beta)}$ as well. We then measure the difference in the AUC between these two experiments, determining how much knowledge about layer $\beta$ helps us predict layer $\alpha$. We call this the \emph{two-layer AUC}. Similarly, the \textit{three-layer AUC} tells us how much knowledge about two layers $\beta$ and $\gamma$ help us predict $\alpha$, and so on. Notice that if different layers have independent structure, without common underlying communities, then including one in the training set much actually decreases our ability to predict the other, causing the AUC to go down.

Computing all \textit{$\ell$-layer AUCs} would require us to try ${L \choose \ell}$ subsets of the layers, which becomes computationally expensive as $\ell$ increases. To avoid this computational bottleneck, we use a greedy bottom-up procedure in which we add one layer at a time to the training dataset, whichever one most increases the AUC, until as many layers as desired have been added. 
This allows us to find a small set of layers which together make it possible to predict links accurately. While this greedy procedure is not guaranteed to find the best possible subset of a given size, it is computationally efficient. 

Alternatively, if the goal is to decide which layers are less informative in predicting the others, in order for instance to compress information by discarding less informative layers, we can use a top-down procedure which starts with all $L$ layers, then iteratively removes whichever one decreases the AUC the least, until a small informative subset of layers remains. We did not pursue this here.

In addition to this link prediction approach, we also cluster the affinity matrices $w^{(\al)}$ in a way similar to~\cite{stanley2015}. That is, we treat the inferred $w^{(\al)}$ as $K^2$-dimensional vectors, and cluster them in $K^2$-dimensional space using the $k$-means algorithm. 
Results for both notions of layer interdependence are shown in the following section. 


\begin{table}[htbp]
\begin{tabular}{lcccccccccc} \toprule
 Network & K & {\small MULTITENSOR} & Diagonal & BPTF \\ \hline
Te\underbar npa\d t\d ti Village & 4 & \textbf{0.89}& 0.88 & 0.87\\ 
A\underbar lak\= apuram Vill. & 6 & \textbf{0.93} &	0.92 & 0.91 \\ 
Malaria & 3 & \textbf{0.83} &	0.82 & 0.82 \\
Malaria & 5 & \textbf{0.86}&	0.85 & 0.85 \\ 
Malaria & 8 & 	\textbf{0.88}& \textbf{0.88} & \textbf{0.88} \\ \bottomrule
\end{tabular}
\caption{AUC for link prediction on real networks using our {\small MULTITENSOR} model, the diagonal/PARAFAC algorithm and Bayesian Poisson Tensor Factorization (BPTF)~\cite{schein2015}. Here we look at the entire dataset at once, and define the AUC as the probability a random link is ranked above a random non-link. Results correspond to the maximum likelihood fixed point over 100 runs of each algorithm with random initial conditions. The best performance for each network is indicated by boldface. All three algorithms perform quite well when shown all layers at once, although {\small MULTITENSOR}'s performance is the best by a small margin.}
\label{table:AllAuc}
\end{table}

\begin{figure*}[ht]%
 \centering
 \subfloat{{\includegraphics[width=4cm]{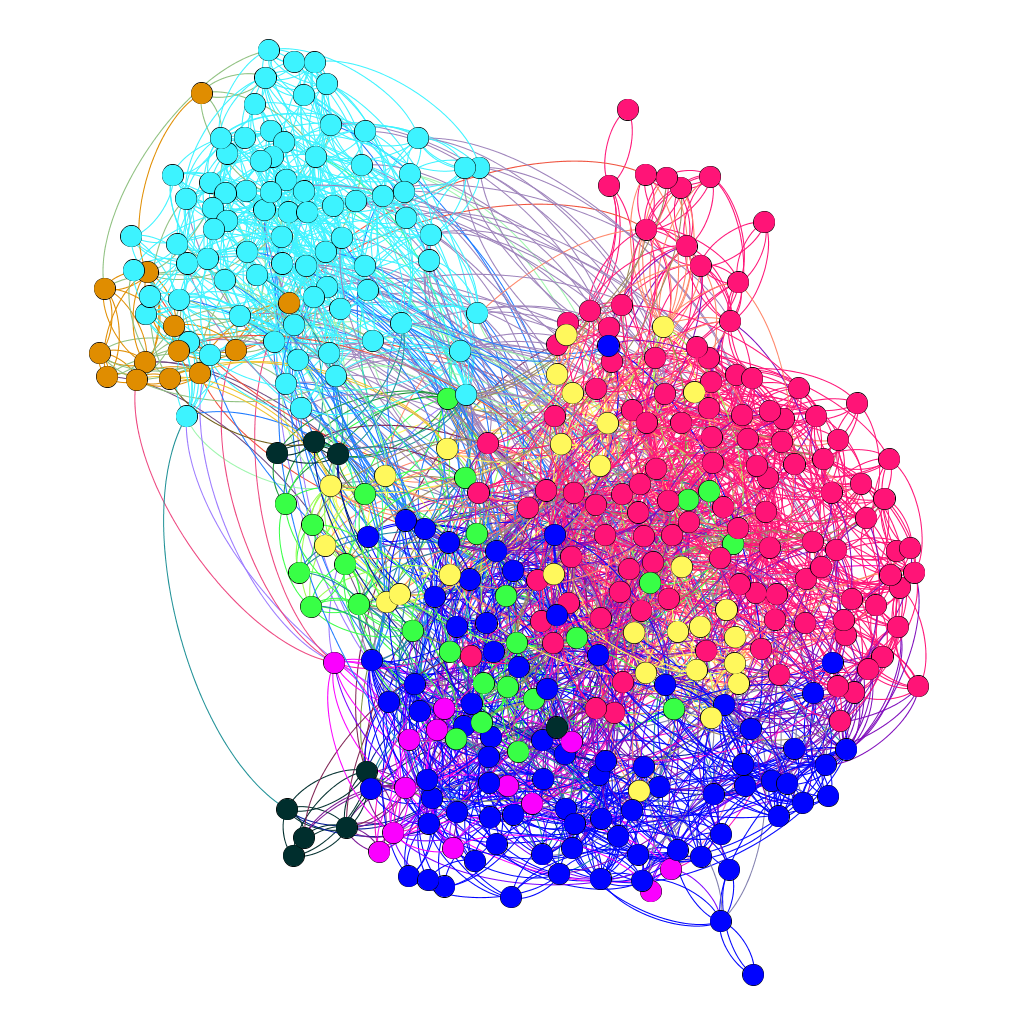} }}%
 \subfloat{{\includegraphics[width=0.19\linewidth]{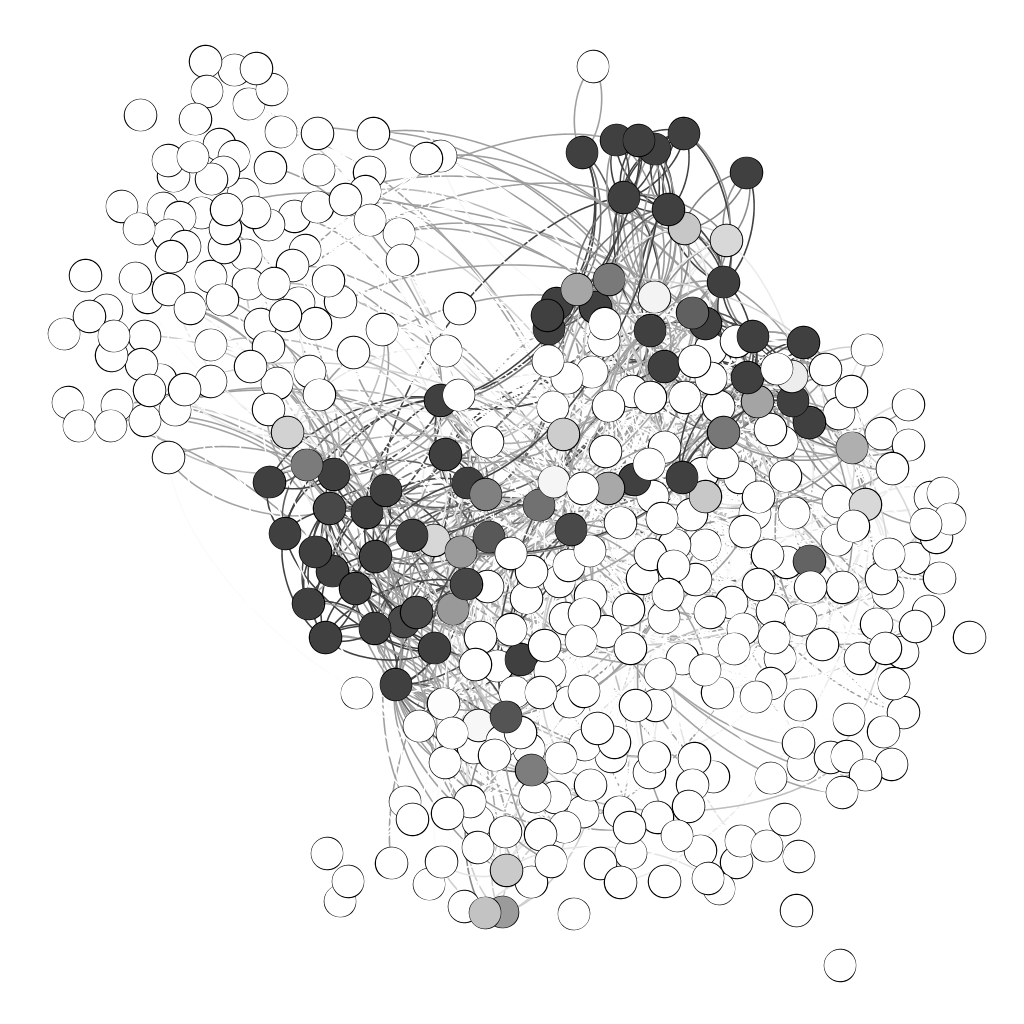} }}%
 \subfloat{{\includegraphics[width=0.19\linewidth]{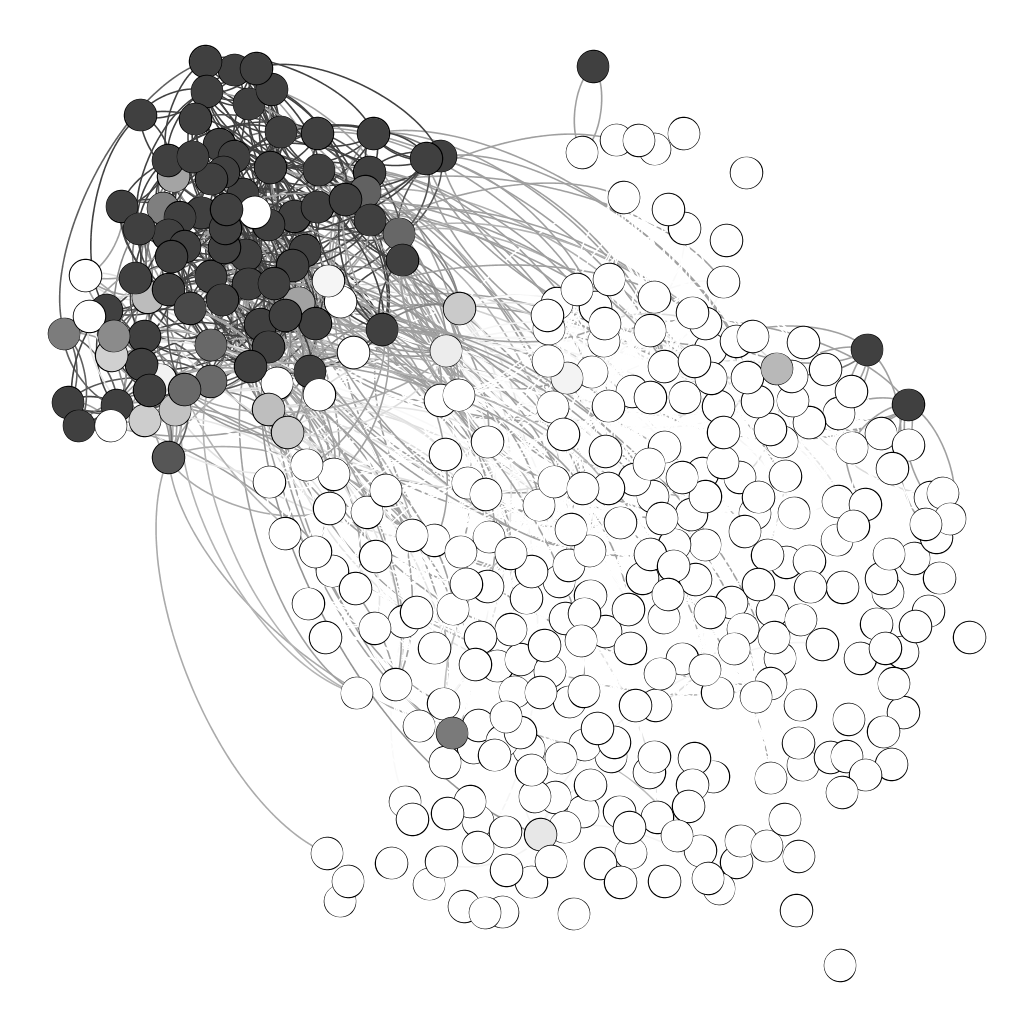} }}%
 \subfloat{{\includegraphics[width=0.19\linewidth]{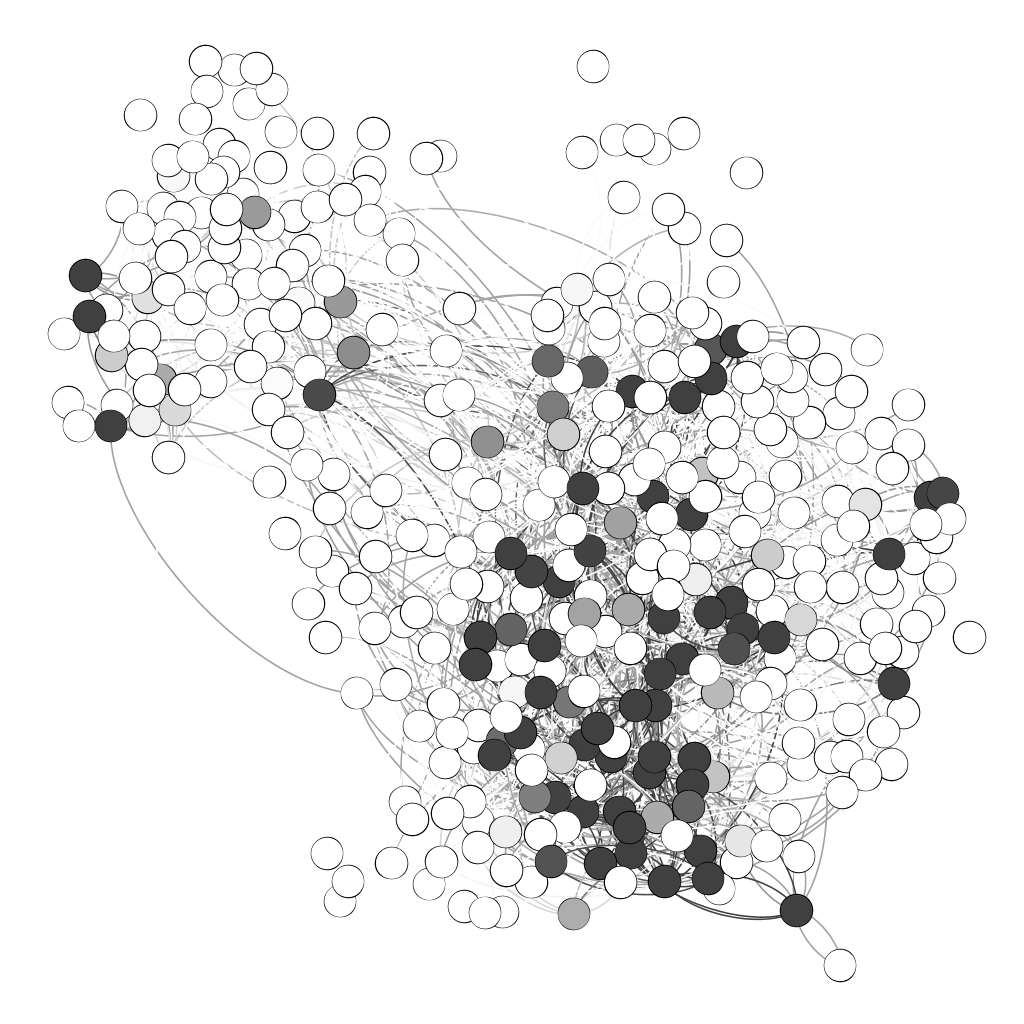} }}%
 \subfloat{{\includegraphics[width=0.19\linewidth]{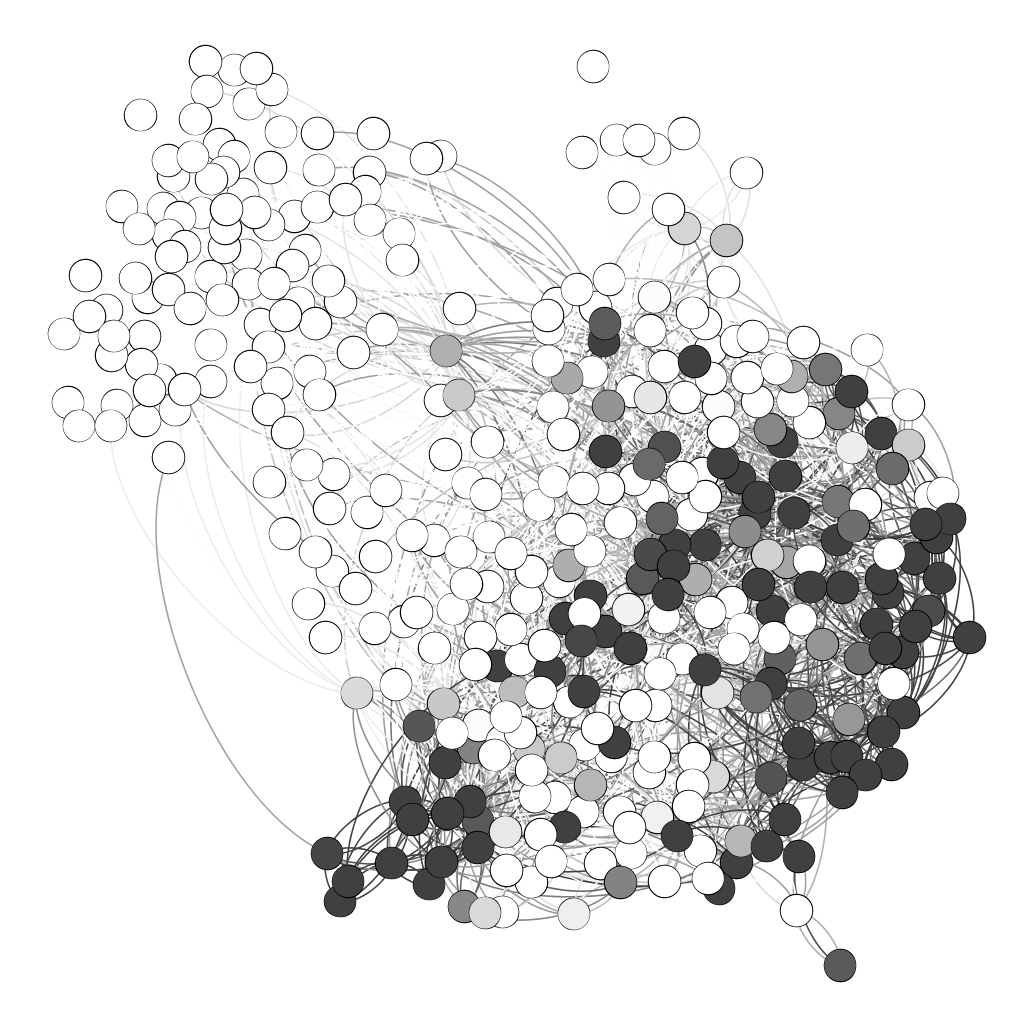} }}%
 \caption{Te\underbar npa\d t\d ti Village community partition. On the left we show the division by caste membership. To the right we show the membership in each of the 4 communities for each node (each figure represents one community), with color ranging from white if the normalized out-going membership $u_{ik}=0$ to black if $u_{ik}=1$. Values in between denote overlapping membership (grey). The fact that caste membership partially overlaps with the communities identified by our algorithm suggests a relationship between topological structure and caste, a topic that will be investigated in a future paper.} %
 \label{fig:partition1}%
\end{figure*}

\begin{figure}[htbp]
	\centering
	\includegraphics[width=1.0\linewidth]{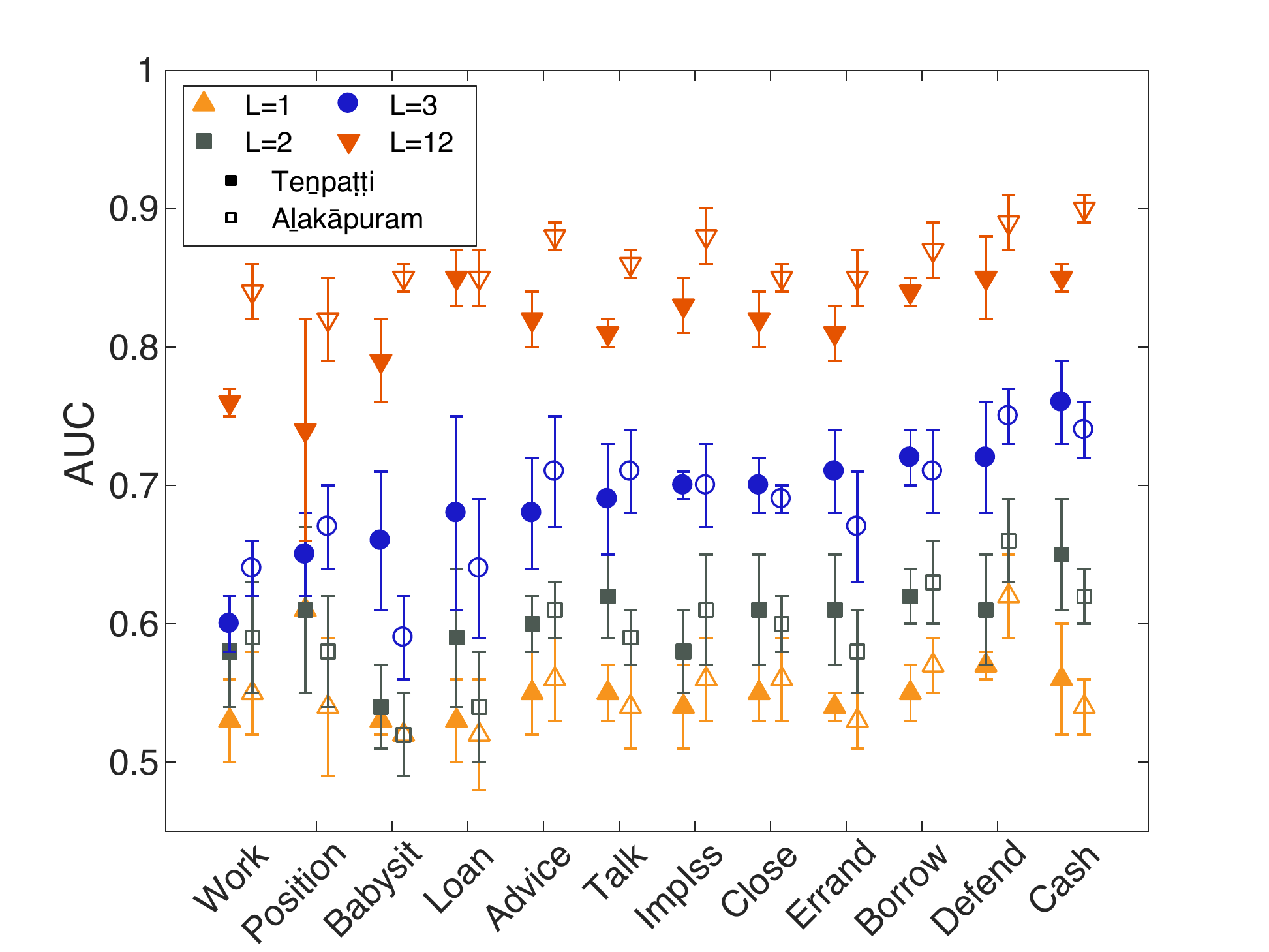}
	\caption{Layer interdependence in the Indian social support networks. On the $x$-axis are the layers' labels used in the test dataset, and the $y$-axis shows the AUC obtained through the cross-validation schemes for measuring layer interdependence. Bold lines are for Te\underbar npa\d t\d ti Village, dashed for A\underbar lak\= apuram Village. $L=1$ refers to single-layer AUC, where the algorithm is only given access to that layer. $L=2, 3, 12$ show the increase in the AUC for that layer when the algorithm is given access to $L$ layers; for $L=2$ and $L=3$ we choose the best set of $L-1$ additional layers using the greedy procedure described in the previous section.}
	\label{fig:ellylayers}
\end{figure}

\begin{figure}[htbp]
	\centering
	\includegraphics[width=\linewidth]{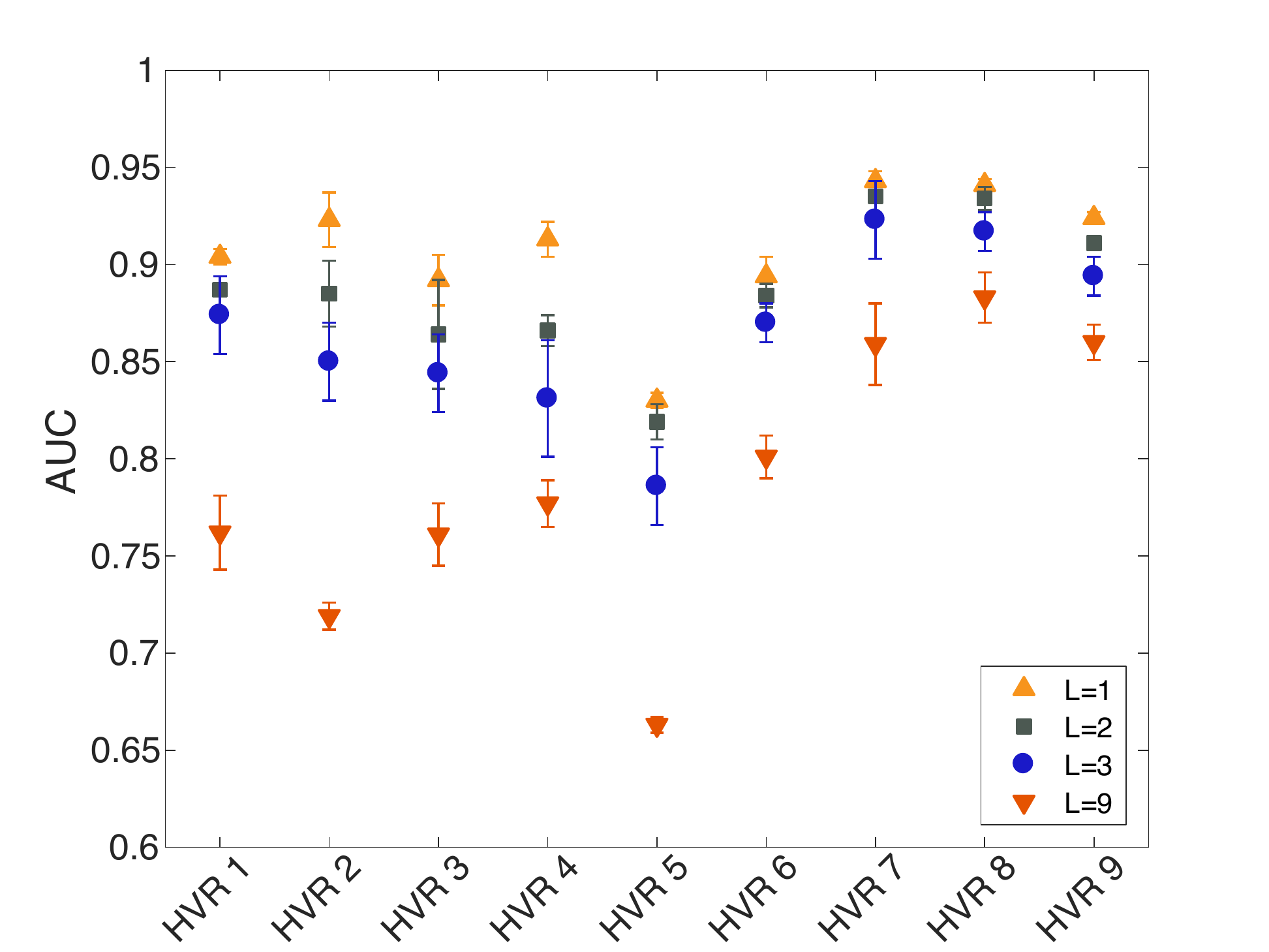}
	\caption{Layer interdependence in the malaria network. Each of the 9 layers corresponds to a so-called ``highly variable region'' (HVR) of the malaria parasite genes, indicated on the $x$-axis, and the $y$-axis shows the AUC obtained through the cross-validation schemes for measuring layer interdependence. $L=1$ refers to single-layer AUC, where the algorithm is only given access to that layer. $L=2, 3, 9$ show the increase in the AUC for that layer when the algorithm is given access to $L$ layers; for $L=2$ and $L=3$ we choose the best set of $L-1$ additional layers using the greedy procedure described in the previous section. Points and error bars are the average and standard deviation over the $5$ folds of cross-validation. Unlike the social support networks, we see that the accuracy of predicting one layer actually decreases when we include others in the training set, indicating that the different layers have independent structure.}
	\label{fig:malarialayers}
\end{figure}

\begin{figure}[htbp]%
	\centering
	\includegraphics[width=0.48\linewidth]{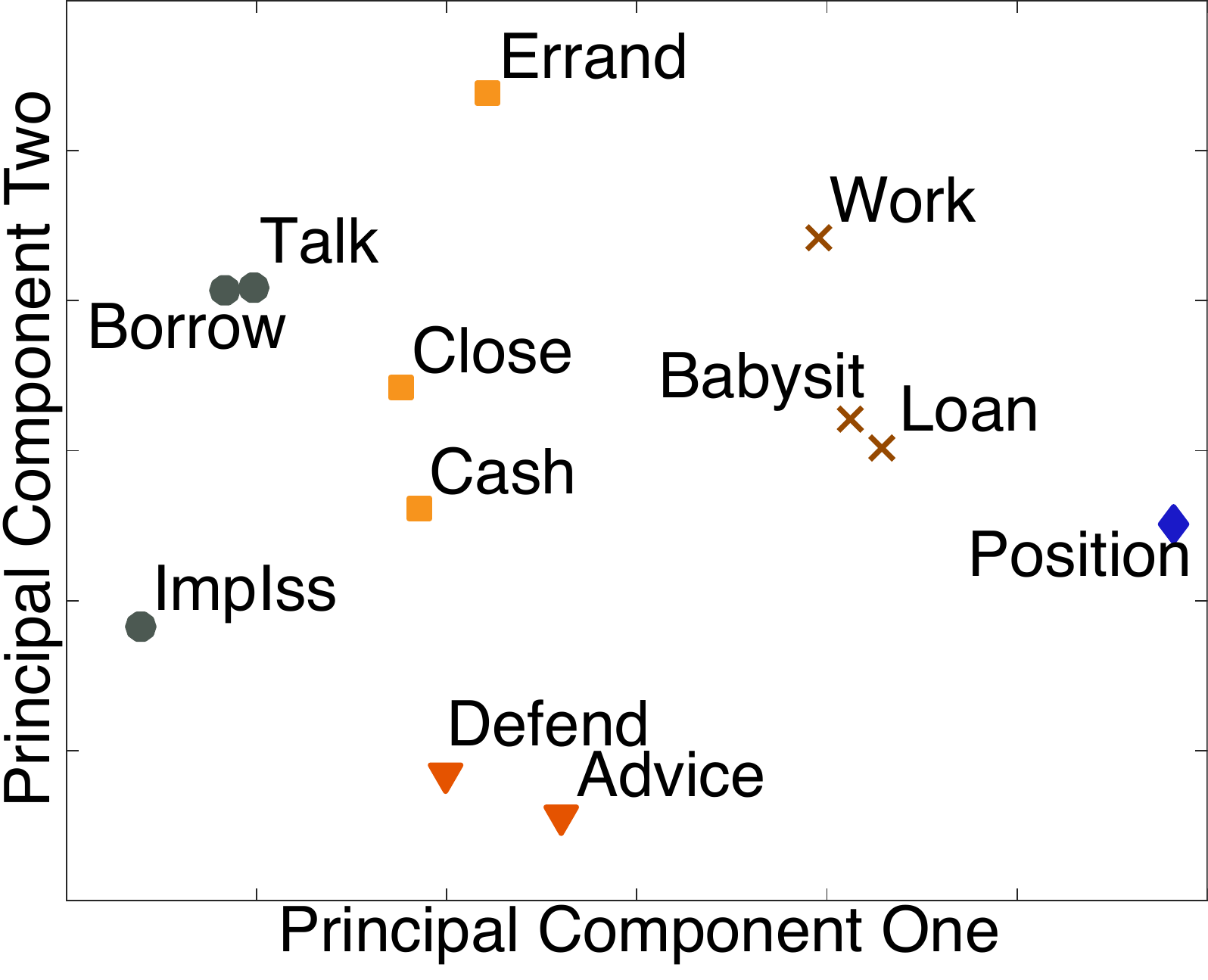}
	\includegraphics[width=0.48\linewidth]{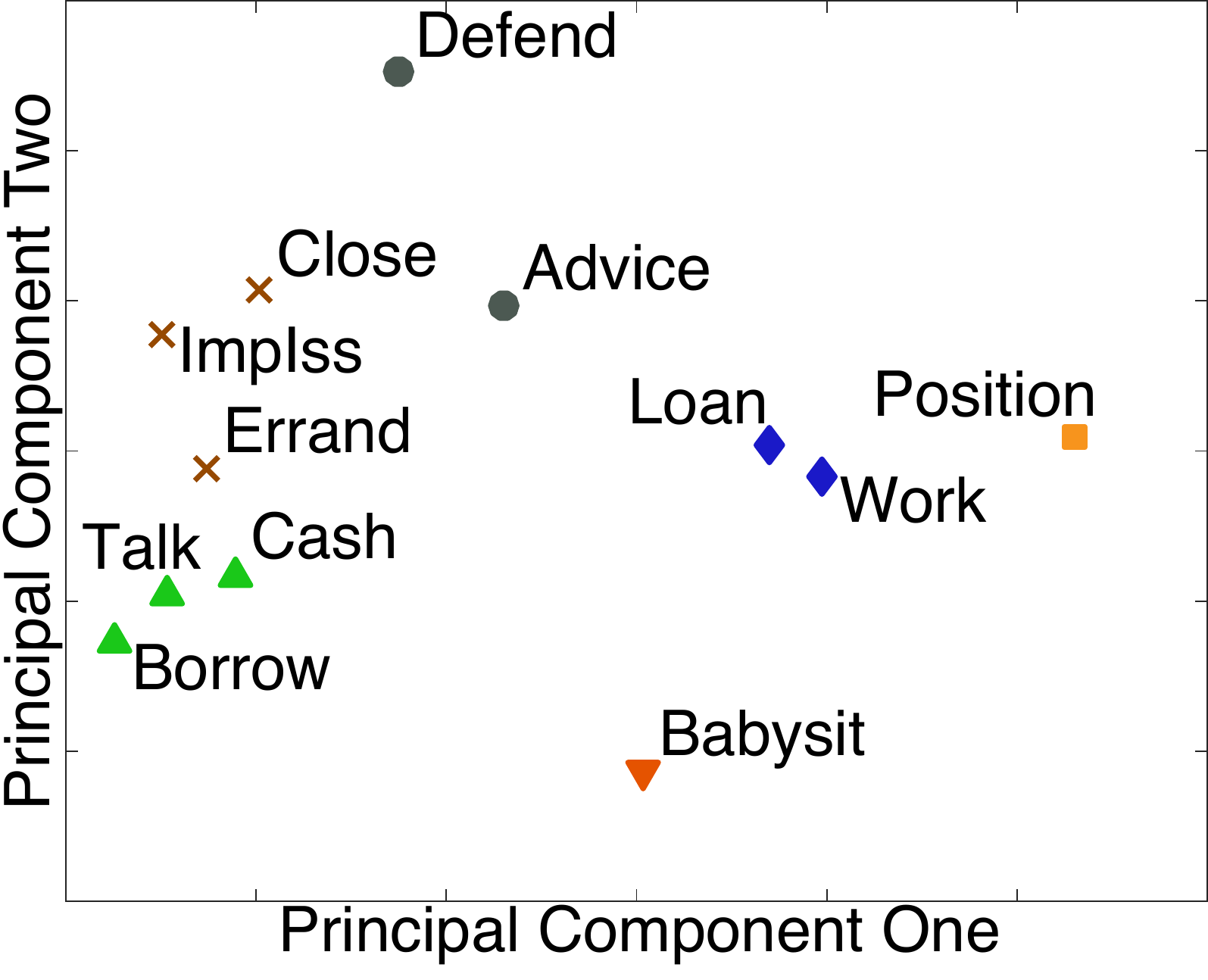}
	\caption{Clusters of the affinity matrices in the layers of the Indian village networks, for Te\underbar npa\d t\d ti on the left and A\underbar lak\= apuram on the right. Cluster labels were obtained using the $k$-means algorithm, treating each $w^{(\al)}$ as a $K^2$-dimensional vector, and we use PCA to visualize them in two dimensions.}
	\label{fig:ellypca}
\end{figure}

\section{Link prediction and layer interdependence in real networks}
\label{sec:results}

To demonstrate our {\small MULTITENSOR} model and algorithm beyond synthetic data, we apply it to two real-world multilayer networks. 
In one network, the {\small MULTITENSOR} model finds that many layers are interdependent, revealing a shared community structure among them. However, in the other network, the models finds the layers to be independent, concluding that there exists no shared structure among them. Together, these two different scenarios illustrate the concrete use of our method in both positive result and negative result scenarios, both of which are likely to arise when analyzing real-world data.

First, we analyze social support networks from two villages in the Indian state of Tamil Nadu, which we call by the pseudonyms~\footnote{Pseudonyms for villages are used for privacy reasons.} ``Te\underbar npa\d t\d ti'' and ``A\underbar lak\= apuram''~\cite{power2015, power2017}. As part of a survey questionnaire, village residents were asked to name those individuals who provided them with 12 different types of support, ranging from lending them household items to helping them navigate government bureaucracy. The resulting directed networks have $N=362$ and $N=420$ nodes, respectively. Each type of support corresponds to a layer in these networks, giving each of them $L=12$ layers, with average degrees ranging from $2.0$ to $4.4$.

Second, we analyze the patterns of shared genetic substrings among a set of malaria parasite virulence genes~\cite{larremore2013}. Each of the $N=307$ nodes represents a single gene, and an edge connects two genes if they share a substring of significant length. Due to the fact that the same set of genes was analyzed at nine different genetic loci (i.e., locations on the genes themselves) which are called ``highly variable regions'' (HVRs), this undirected network has $L=9$ layers, with average degrees ranging from $5.1$ to $76.4$. 

The scientifically interesting questions for both networks revolve around the mechanisms driving edge formation. Hypothesized factors include kinship and caste in the Indian social support networks, and upstream promotor sequence or parasite origin in the malaria genetic networks. However, addressing these questions is beyond the scope of this paper, where we instead wish to evaluate the effectiveness of our algorithm. 

One option would be to use our algorithm to cluster the nodes, and compare the resulting group assignments with metadata such as gender, caste, or geographical location. Indeed, in Figure~\ref{fig:partition1} we show the community assignment for Te\underbar npa\d t\d ti predicted by our model, and compare it with the division of individuals into castes. Although the figure suggests that the partition might be correlated with caste membership, we do not expect this to be the only type of metadata correlated with the community structure, and we do not consider this correlation to be a good measure of accuracy. 
Here we focus instead on link prediction, and in particular on the extent to which knowledge of some layers helps us predict links in others, as described in the previous section.

As for the synthetic networks, our {\small MULTITENSOR} algorithm, the Diagonal/PARAFAC algorithm, and the BPTF algorithm each provide a framework for link prediction. Table~\ref{table:AllAuc} reports the AUC over each entire network and for each algorithm. The algorithms' performance are roughly similar, although our algorithm has slightly higher performance. This suggests that these networks are primarily assortative; this is certainly true of the malaria network, since it is defined in terms of similarity.

To measure layer interdependence we implemented the method described in the previous section, where we attempt to predict the adjacency matrix of a given layer with 20\% of its entries held out, and give the algorithm access to a subset of other layers as part of its training dataset. Interestingly, we obtain opposite results in these two cases. 

For the social networks, we find that increasing the number of layers in the training dataset does indeed improve link prediction, with a performance that increases monotonically with the number of additional layers. In Figure~\ref{fig:ellylayers} we show that the AUC for each layer as a function of the number of layers the algorithm is given access to. We found that the best number of groups for link prediction was $K=4$ for the first village and $K=6$ for the second one.

Many layers viewed on their own ($L=1$) are difficult to predict, with AUCs just above $0.5$, i.e., only slightly better than chance. By giving the algorithm access to one more layer ($L=2$) the AUC typically improves by only about $0.05$. However, if we give it access to two additional layers ($L=3$) the AUC improves significantly for almost all of the layers, and this is even more true when we give it access to the entire dataset. (For $L=2, 3$ we use the greedy procedure to choose which $L-1$ layers to add to the training dataset.) 

Thus in these social networks, the {\small MULTITENSOR} algorithm is able to usefully apply knowledge from some layers to others. Interestingly, we also see consistency between the two villages with regard to which layers are the hardest to predict, and which layers are the most helpful to include in the training dataset. In particular, the ImpIss layer (``Who do you discuss important matters with?'') is helpful in predicting many layers, while Position, Work, Loan, and Babysit are much less so, and in some cases even decrease the AUC.

We can compare this with the clustering of the $L$ affinity matrices we obtained using standard clustering algorithms, in a spirit similar to~\cite{stanley2015}. In Figure~\ref{fig:ellypca} we use Principal Component Analysis~\cite{jolliffe2002} to visualize the $L$ matrices $w^{(\al)}$, projecting them along two principal directions in $K^2$-dimensional space, and we give them cluster labels using the $k$-means algorithm~\cite{macqueen1967}. Indeed we see that Position, Work, Loan, and Babysit are farther from the others, suggesting that these layers are structurally quite different from the others; note also in Figure~\ref{fig:ellylayers} that these layers are among the hardest to predict. In contrast, ImpIss is closer to the other layers, at least for the second village, consistent with the fact that it often helps predict other layers. We also find for $L=2$, that the Borrow layer is the most helpful when predicting the Talk layer in both villages, which is consistent with the fact that these two layers are clustered close together.

In contrast, for the malaria network we find that the best performance is obtained when no other layer is added to the dataset, meaning that prediction actually worsens monotonically as we increase the number of added layers, as shown in Figure~\ref{fig:malarialayers}. This seems to corroborate past findings~\cite{larremore2013} in an important way. Specifically, the standing hypothesis about these genes is that they are maximally diverse in order to most effectively evade the immune system. If there were correlations between loci, which we would see here as the ability of one layer to help in the link prediction of another layer, then this would diminish these genes' overall diversity. This would diminish the amount of ``immune evasion space'' that is spanned by the parasites, and would therefore result in an overall fitness decrease for the parasites.

\section{Conclusions}
\label{sec:conclusion}

We have proposed a generative model for multilayer networks that extends and generalizes the mixed-membership stochastic block model. It assumes that the layers share a common community structure, but allows links in different layers to be correlated with the community memberships in different ways, such as assortative, disassortative, core-periphery, or hierarchical structure, or arbitrary mixtures thereof.  It explicitly allows the communities to overlap, and can be applied to networks with directed, undirected, or integer-weighted links.  We showed that it can be fit to large datasets using a scalable expectation-maximization algorithm, whose running time per iteration is linear in the total size of the dataset and which converges quickly in practice. Due to its ability to describe a wide variety of graph structures, it performs well on synthetic and real data, in terms of both community detection and link prediction.

In addition to performing community detection, the methods in this paper naturally incorporate a framework for link prediction, which we use as a quantitative definition of interdependency between the network's layers.  Namely, we measure how much knowledge of one layer, or a set of layers, improves the accuracy of link prediction in another layer.  This measure is quite general, and goes beyond approaches that cluster layers into strata with similar parameters (e.g.~\cite{stanley2015}); for instance, if two layers both depend strongly on the underlying communities, they will be interdependent in this sense even if one is assortative and the other is disassortative, making their affinity matrices very different.  In addition to providing hints about causal or structural relationships between the layers, this notion of interdependence may be useful to choosing weights for multilayer versions of common network measures, such as eigenvector centrality~\cite{sola2013} and modularity~\cite{mucha2010}. The same link prediction and cross-validation framework used to quantify layer interdependence can also be used to identify and avoid overfitting.

Beyond establishing high performance on synthetic datasets, we also applied our methods to two real-world datasets. We found patterns of interdependence between layers of social networks from two Indian villages, indicating correlations between different kinds of social ties, and confirmed that these patterns are largely consistent between the two villages. In contrast, when we applied our methods to a multilayer network of sequence sharing among malaria's virulence genes, we found that the layers were essentially unrelated. This suggests that similarities at different loci of the amino acid sequences are evolving under uncorrelated constraints, rigorously confirming a result based on independent analyses of each layer~\cite{larremore2013}. In both cases, our {\small MULTITENSOR} approach provided information not revealed in previous studies of these datasets, and proved to be useful in identifying not only the presence of meaningful structure, but its absence as well.

The solution we provide for the layer interdependence problem may find application beyond the analysis of extant datasets. Because our method can be used to aggregate layers into clusters, or to compress a dataset by identifying especially relevant or redundant layers, it can direct experimentalists or field researchers in learning which data to collect or prioritize. For example, if two layers of a social network are found by our methods to be redundant during a pilot study, the redundant layer need not be collected at scale. Particularly in cases where data collection is labor intensive, expensive, or generally difficult, robust solutions to the layer interdependence problem can help maximize the impact of studies constrained by limited resources in the laboratory or the field. On the other hand, when layers are found to be independent of each other, our methods provide justification for comprehensive data collection of the relevant layers. 

\section*{Acknowledgements}
This work was supported by the John Templeton Foundation (CDB, CM), the Army Research Office under grant W911NF-12-R-0012 (CM), a National Science Foundation Doctoral Dissertation Improvement grant (BCS-1121326), a Fulbright-Nehru Student Researcher Award, the Stanford Center for South Asia, and Stanford University (EAP), and the Santa Fe Institute Omidyar Fellowship (DBL, EAP). We are grateful to Hanna Wallach and Aaron Schein for helpful discussions.

\bibliographystyle{apsrev4-1}
\bibliography{bibliography}

\end{document}